\documentclass[twocolumn,showpacs,preprintnumbers,aps,amsmath,amssymb]{revtex4}

\usepackage[dvips]{graphicx}

\begin{document}


\title{Experimental Investigation of Nonideal Two-qubit Quantum-state Filter
by Quantum Process Tomography}
\author{Yoshihiro Nambu}
\author{Kazuo Nakamura}
\affiliation{Fundamental and Environmental Research Laboratories, NEC, 34 Miyukigaoka,
Tsukuba, Ibaraki 305-8501, Japan}

\begin{abstract}
We used quantum process tomography to investigate and identify the function
of a nonideal two-qubit quantum-state filters subject to various degree of
decoherence. We present a simple decoherence model that explains the
experimental results and point out that a beamsplitter followed by a
post-selection process is not, as commonly believed, a singlet-state filter.
In the ideal case it is a triplet-state filter.
\end{abstract}

\pacs{03.65.Wj, 42.50.Dv}
\keywords{Characterization, operation, decoherence}

\maketitle

Characterization of quantum devices is important, and the characterization
of single- and two-qubit devices is particularly important because a
single-qubit rotation gate and a two-qubit controlled-NOT (C-NOT) gate are
two building blocks of the quantum computer\cite{Elementary gate}. The
quantum systems in actual devices may undergo nonunitary state evolution
because they interact with their environments, so identifying and
understanding the function of such nonideal devices are of practical
importance. Nonideal devices can be characterized by quantum process
tomography (QPT)\cite{QPT Theory1,QPT Theory2,QPT Theory3}, and this method
has been used to characterize single-qubit transmission channels\cite%
{Single-qubit QPT1,Single-qubit QPT2} and rotation gates for photons\cite%
{Single-qubit QPT3}, gates for ensembles of two-qubit systems in NMR\cite%
{Two-qubit QPT1}, and a two-qubit quantum-state filter\cite{Two-qubit QPT2}
and a C-NOT gate for photons\cite{Two-qubit QPT3}.

This letter reports the QPT of a quantum-state filter---the beamsplitter
(BS) in a Hong-Ou-Mandel (HOM) interferometer\cite{HOM,Dense coding}---as a
demonstration how the function of two-qubit devices can be identified
experimentally. The result of an earlier investigation of the nearly ideal
filter\cite{Two-qubit QPT2} was inconsistent with a common belief that an
ideal BS followed by a post-selection process---only the outcomes where the
two input photons come out in two output ports are taken into
account---behaves as a singlet-state filter \cite{Bell Measurement,Dense
coding}, but the reason for this inconsistency was not fully understood. To
resolve this problem, we have investigated the function of the nonideal
filters subject to various degree of decoherence and have identified two
leading contributions to the operator-sum representation of the quantum
operation for them\cite{Elementary gate,QPT Theory1,CP map}. Here we present
a model that explains our experimental results and resolve the above
inconsistency.

Let us start by introducing the process matrix\cite{Two-qubit QPT3}, which
offers us the most general description of the quantum device allowed in
quantum mechanics, and showing how to obtain it experimentally. Consider a
two-qubit device whose function is to map a two-qubit input density operator 
$\hat{\rho}_{in}$ to a two-qubit output density operator $\hat{\rho}_{out}=%
\mathcal{E}(\hat{\rho}_{in})$, where $\mathcal{E}$ is a completely positive
linear map\cite{CP map}. The map $\mathcal{E}$ is considered a superoperator
acting on positive operators\cite{Super operator}. Let us introduce the set
of standard operator basis $\{\hat{X}_{k}\}_{k=0}^{3}$ that spans two-qubit
operator space\cite{Operator basis}, where $\hat{X}_{\langle ij\rangle
}=\vert i\rangle\langle j\vert $ ($i,j=0,1$) and $\left\langle
ij\right\rangle :=2i+j$. Then $\mathcal{E}$ can be expanded as 
\begin{equation}
\mathcal{E}(\hat{\rho})={\displaystyle\sum\limits_{\substack{ k,l,m,n=0}}^{3}%
}\left( \chi^{S}\right) _{[kl][mn]}\hat{X}_{k}\otimes\hat{X}_{l}\hat{\rho }%
\hat{X}_{m}^{\dagger}\otimes\hat{X}_{n}^{\dagger},  \label{eq1}
\end{equation}
where $[kl]:=4k+l$\cite{Super operator}. A 16$\times$16 positive Hermitian
matrix $\chi^{S}$ fully characterizes the function of quantum device and is
called the process matrix in the standard operator basis\cite{Two-qubit QPT3}%
.

We can transform $\chi^{S}$ into the process matrix in any operator basis
set $\{\hat{A}_{\alpha}\}_{\alpha=0}^{15}$ that spans the two-qubit operator
space. Note that $\hat{A}_{\alpha}$ may be defined either locally or
nonlocally. For example, the Kronecker products of the Pauli operator basis $%
\hat{B}_{[kl]}=\hat{E}_{k}\otimes\hat{E}_{l}$ ($k,l=0,\cdots,3$), where $\{%
\hat{E}_{k}\}$=$\{\hat{\sigma}_{0}/\sqrt{2},\hat{\sigma}_{1}/\sqrt{2},\hat{%
\sigma}_{2}/\sqrt{2},\hat{\sigma}_{3}/\sqrt{2}\}$ and $\hat{\sigma}_{0}$
denotes an identity operator, is a local operator basis, whereas the outer
product $\hat{C}_{[kl]}=\left\vert \Phi_{k}\right\rangle \left\langle \Phi
_{l}\right\vert $, where $\left\vert \Phi_{k}\right\rangle $ are Bell
states, is a nonlocal one. The operator basis $\hat{A}_{\alpha}$ is, in
general, unitarily related to the standard operator basis $\hat{X}_{k}\otimes%
\hat {X}_{l}$. That is, $\hat{A}_{\alpha}={\textstyle\sum%
\nolimits_{k,l=0}^{3}}\left( U\right) _{[kl]\alpha}\hat{X}_{k}\otimes\hat{X}%
_{l}$, where \textit{U} is a 16$\times$16 unitary matrix. Then $\mathcal{E}$
can be expanded using $\hat{A}_{\alpha}$ as follows: 
\begin{equation}
\mathcal{E}(\hat{\rho})={\displaystyle\sum\limits_{\alpha,\beta=0}^{15}}%
\left( \chi^{A}\right) _{\alpha\beta}\hat{A}_{\alpha}\hat{\rho}\hat {A}%
_{\beta}^{\dagger},  \label{eq2}
\end{equation}
where $\chi^{A}=U^{\dagger}\chi^{S}U$. Therefore, $\chi^{A}$ is given by the
unitary transform of $\chi^{S}$.

To obtain $\chi ^{S}$, it is sufficient to determine the associated
four-qubit state\cite{four-qubit state}. Let us consider a four-qubit system
composed of the two two-qubit systems 1-2 and 3-4 and introduce the
double-ket notation $\left\vert I\right\rangle \rangle :=(\hat{I}\otimes 
\hat{I})\left\vert \Phi \right\rangle $ to denote the bipartite state, where 
$\left\vert \Phi \right\rangle ={\textstyle\sum\nolimits_{i=0}^{1}}%
\left\vert i,i\right\rangle $ is the unnormalized maximally entangled state%
\cite{QPT Theory3,Operator basis}. Let us introduce the unnormalized
four-qubit density operator $\hat{\mathcal{D}}_{\mathcal{E}}$ associated
with $\mathcal{E}$ by $\hat{\mathcal{D}}_{\mathcal{E}}=\mathcal{E}%
^{(13)}\otimes \mathcal{I}^{(24)}(\left\vert I\right\rangle \rangle
_{12}\langle \left\langle I\right\vert \otimes \left\vert I\right\rangle
\rangle _{34}\langle \left\langle I\right\vert )$, where $\mathcal{I}$ is an
identity map and the superscript and subscript respectively denote the space
on which each map acts and the space to which the operator belongs\cite%
{four-qubit state}. Using Eq. \ref{eq1} and $|X_{k}\rangle\rangle :=(\hat{X}%
_{k}\otimes \hat{I})|\Phi \rangle $, we can rewrite $\hat{\mathcal{D}}_{%
\mathcal{E}}$ as 
\begin{equation}
\hat{\mathcal{D}}_{\mathcal{E}}={\displaystyle\sum\limits_{\substack{ %
k,l,m,n=0}}^{3}}\left( \chi ^{S}\right) _{[kl][mn]}|X_{k}\rangle \rangle
|X_{l}\rangle \rangle \langle \langle X_{m}|\langle \langle X_{n}|.
\label{eq3}
\end{equation}%
If we note that $|X_{\langle ij\rangle }\rangle \rangle \equiv |i,j\rangle $%
, Eq. \ref{eq3} indicates that $\chi ^{S}$ agrees exactly with the density
matrix of the associated state $\mathcal{\hat{D}}_{\mathcal{E}}$ given in
the standard state basis. Evaluation of the process matrix is thus reduced
to evaluation of the associated state $\mathcal{\hat{D}}_{\mathcal{E}}$.

To obtain $\hat{\mathcal{D}}_{\mathcal{E}}$, it is sufficient to evaluate
all the maps $\mathcal{E}(\hat{X}_{\langle ij\rangle }\otimes \hat{X}%
_{\langle kl\rangle })$ for all the elements of the standard operator basis $%
\hat{X}_{\langle ij\rangle }\otimes \hat{X}_{\langle kl\rangle }$, since we
can obtain the permutation equivalent of $\hat{\mathcal{D}}_{\mathcal{E}}$%
---that is, $\mathcal{\tilde{D}}_{\mathcal{E}}=\hat{P}_{23}\hat{P}_{12}\hat{P%
}_{34}\mathcal{\hat{D}}_{\mathcal{E}}\hat{P}_{34}\hat{P}_{12}\hat{P}_{23}=%
\mathcal{I}^{(12)}\otimes \mathcal{E}^{(34)}(|I\rangle \rangle _{13}\langle
\langle I|\otimes |I\rangle \rangle _{24}\langle \langle I|)$---by using the
following identity: 
\begin{equation}
\mathcal{\tilde{D}}_{\mathcal{E}}\equiv {\displaystyle\sum\limits_{k,l=0}^{3}%
}\hat{X}_{k}\otimes \hat{X}_{l}\otimes \mathcal{E}(\hat{X}_{k}\otimes \hat{X}%
_{l}),  \label{eq4}
\end{equation}%
where $\hat{P}_{ij}$ is a permutation operator between the systems \textit{i}
and \textit{j}. To evaluate these maps, we need only to use state tomography%
\cite{State tomography} to evaluate the 16 two-qubit output states $\{%
\mathcal{E}(\hat{\rho}_{i}^{(1)}\otimes \hat{\rho}_{j}^{(2)})\}_{i,j=0}^{3}$
associated with the set of 16 separable input states, where $\hat{\rho}%
_{i}\in \{ \vert 0\rangle \langle 0\vert ,\vert 1\rangle \langle 1\vert
,\vert +\rangle \langle +\vert ,\left\vert\circlearrowleft \rangle \langle
\circlearrowleft \right\vert \} $, $\vert +\rangle =1/\sqrt{2}( \vert
0\rangle +\vert 1\rangle ) $, and $\left\vert \circlearrowleft \right\rangle
=1/\sqrt{2}( \vert 0\rangle+i\vert 1\rangle ) $ \cite{QPT Theory1,QPT
Theory2,QPT Theory3}. We therefore can obtain the process matrix directly
from the experimental data without using a matrix inversion procedure\cite%
{Elementary gate,QPT Theory1}.

We used the above procedure to obtain the process matrices for nonideal
two-qubit quantum-state filters. The experimental setup is shown in Fig. \ref%
{F1}. A 0.3-mm-thick type-I crystal (BBO) was pumped by frequency-tripled
pulses generated by 100-fs pulses from a mode-locked Ti:sapphire laser
(repetition rate $\approx $82 MHz). Pairs of horizontally polarized 532-nm
photons were generated and those traveling in two directions 9${{}^{\circ }}$
from the pump beam were selected by 5-mm irises (1 m from the crystal) and
identical 3-nm-width filters (IF) placed before the detectors. They were
injected into a HOM interferometer\cite{HOM,Dense coding,Two-qubit QPT2}
having a broadband non-polarizing cube BS (Suruga Seiki model S322-20-550).
The horizontally (\textit{H}) and vertically (\textit{V}) polarized states
(respectively $\left\vert 0\right\rangle $ and $\left\vert 1\right\rangle $)
and the other two states required for the QPT were prepared by half- and
quarter-wave plates (HWP and QWP) placed immediately before the BS.
Down-converted photons were incident on the BS at 45${{}^{\circ }}$. After
the BS, polarization analyzers consisting of a QWP and a HWP followed by a
Glan-Thompson prism (POL) and a single-photon counting module (PMT)
performed projective measurements onto the set of the 16 product states $\{%
\hat{\rho}_{i}^{(1)}\otimes \hat{\rho}_{j}^{(2)}\}_{i,j=0}^{3}$. Only
signals coincident at the two PMTs were recorded, and the associated 16
output states $\{\mathcal{E}(\hat{\rho}_{i}^{(1)}\otimes \hat{\rho}%
_{j}^{(2)})\}_{i,j=0}^{3}$ were evaluated by state tomography\cite{State
tomography}. Note that we used unnormalized output density matrices because
the state filtering is trace-decreasing process whose output intensity
depends on the input state. 
\begin{figure}[tbph]
\includegraphics[scale=0.38]{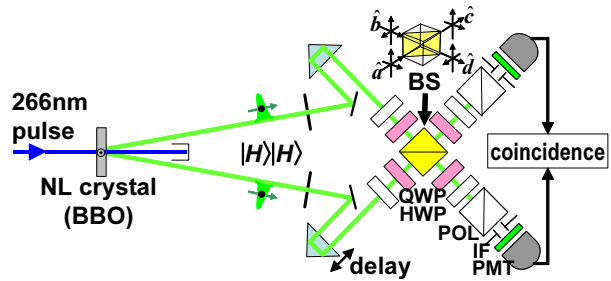}
\caption{Experimental setup to investigate the beamsplitter.}
\label{F1}
\end{figure}

\begin{figure}[tbph]
\includegraphics[scale=0.62]{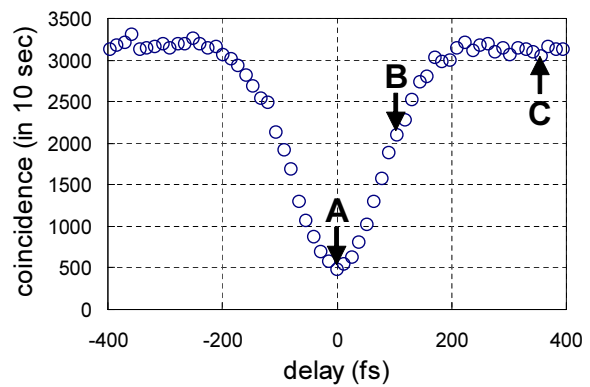}
\caption{HOM dip observed in the coincidence record. Arrows indicate delays $%
\protect\tau$ of (A) 0, (B) 100, and (C) 350 fs introduced in one of the two
arms.}
\label{F2}
\end{figure}

We used optical trombones to equalize the two path lengths to within the
coherence length $l_{c}$ of the down-converted photons ($\approx 200$ $\mu m$%
). Two two-photon amplitudes contributing to coincidence detection---\textit{%
i.e.},that both photons are reflected (\textit{r}-\textit{r}) or that both
are transmitted (\textit{t}-\textit{t})--- were initially ($\tau =0$)
aligned to overlap both spatially and temporally, giving a HOM dip
visibility of 85\% as shown in Fig. \ref{F2}. In this case, the BS is
commonly thought to work as a singlet-state filter\cite{Bell
Measurement,Dense coding}. To investigate the nonideal filter, we introduced
a delay $\tau $ in one arm by moving one of the trombone prisms. The larger
the $\tau $ is, the smaller the overlap of the two amplitudes is and the
more the temporal and polarization information of the photon pair is
entangled. We will show later that this results in decoherence if we observe
only the polarization information. 

\begin{figure*}[tbh]
\begin{center}
\includegraphics*[scale=0.68]{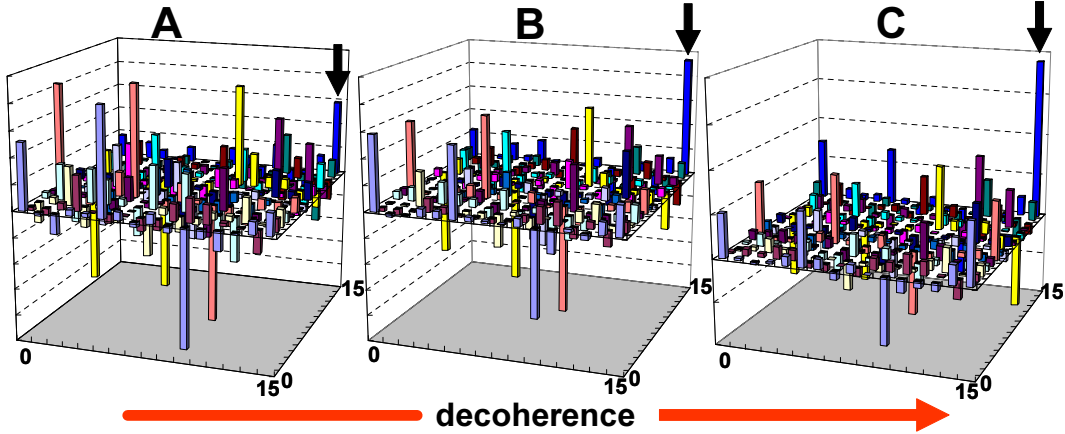}
\end{center}
\caption{ The process matrices of the state filter obtained for delays $%
\protect\tau $ of (A) 0, (B) 100, and (C) 350 fs. Only the real parts of the
matrices are shown (in arbitrary units). Vertical arrows indicate the
diagonal elements $(\protect\chi ^{F})_{15\,15}$.}
\label{F3}
\end{figure*}

We obtained the process matrices $\chi ^{F}$ associated with three-delay
conditions ($\tau =0$, $100$, and $350$ fs). To make the results clear, we
represent them here in the nonlocal operator basis $\hat{F}_{[ij]}=(\hat{E}%
_{i}\otimes \hat{E}_{j})\hat{U}_{S}$, where $\hat{U}_{S}$ is a swap
operator. They are shown in Fig. \ref{F3}, where space is saved by showing
only the real parts. What is most interesting is that the contribution of $%
(\chi ^{F})_{15\,15}$ increases with increasing delay $\tau $.

\begin{figure*}[tbph]
\begin{center}
\includegraphics*[scale=0.68]{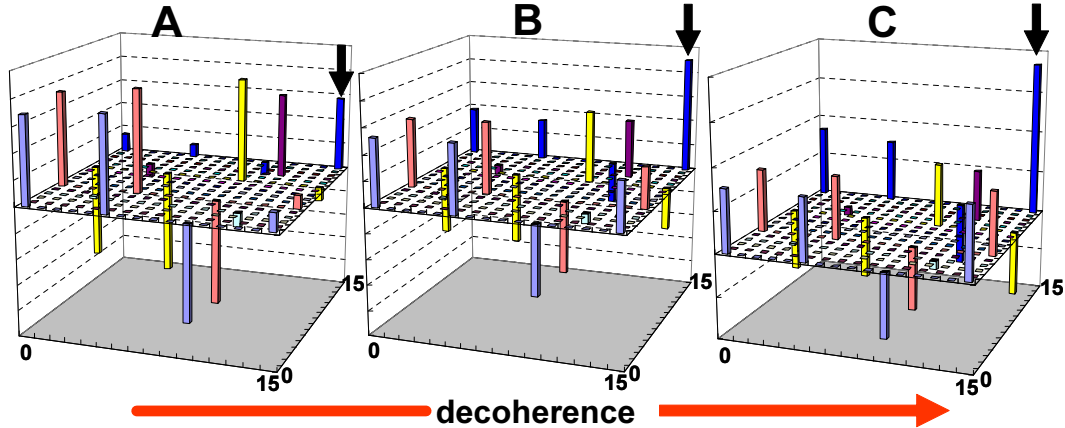}
\end{center}
\caption{ Calculated process matrices corresponding to Fig. \protect\ref{F3}
(see text). The following parameters were used: (A) $p=0.14$, (B) $p=0.325$,
(C) $p=0.5$ and the common parameters $\mathcal{R}/\mathcal{T}=0.76$, $%
\protect\theta _{1}\approx 0.41\protect\pi $, and $\protect\theta %
_{2}\approx 0.076\protect\pi $.}
\label{F4}
\end{figure*}

This result is consistent not with the common belief but with what was
pointed out by Mitchell \textit{et al.}\cite{Two-qubit QPT2}. Indeed, we
found that our results are well reproduced if we assume that $\mathcal{E}(%
\hat{\rho})$ has an operator-sum representation with two Kraus operators $%
\mathcal{\hat{P}}^{-}$ and $\mathcal{\hat{P}}^{+}$\cite{Elementary gate,CP
map}: 
\begin{align}
\mathcal{E}(\hat{\rho})& =(1-p)\mathcal{\hat{P}}^{-}\hat{\rho}\mathcal{\hat{P%
}}^{-\dag }+p\mathcal{\hat{P}}^{+}\hat{\rho}\mathcal{\hat{P}}^{+\dag },
\label{eq5} \\
\mathcal{\hat{P}}^{\pm }& =\mathcal{T}U_{I}\mathcal{\pm R}\hat{U}_{3}(\theta
_{1},\theta _{2})\hat{U}_{S},  \label{eq6}
\end{align}%
where \textit{p} ($0\leq p\leq 1/2$) is a parameter that can be interpreted
as the degree of decoherence and where $\hat{U}_{3}(\theta _{1},\theta
_{2})=e^{i\theta _{1}\hat{\sigma}_{3}/2}\hat{\sigma}_{3}\otimes e^{-i\theta
_{2}\hat{\sigma}_{3}/2}\hat{\sigma}_{3}$, which explains the large
contribution of $(\chi ^{F})_{15\,\ 15}$ instead of $(\chi ^{F})_{0\,0}$ as
would be expected for the nonideal singlet-state filter. The Kraus operators 
$\mathcal{\hat{P}}^{-}$ and $\mathcal{\hat{P}}^{+}$ correspond respectively
to the operator for the ideal state-filter and the noise operator. Note that
Kraus operators are not unique and are not necessarily orthogonal\cite%
{Elementary gate}. Figure \ref{F4} shows the process matrices calculated
using the measured value $\mathcal{R}/\mathcal{T}=0.76$ and appropriate
values of the parameters $\theta _{1}$, $\theta _{2}$, and $p$. As shown in
Fig. \ref{F4}, Eqs. \ref{eq5} and \ref{eq6} well reproduced the experimental
results. It follows that the operator $\mathcal{\hat{P}}^{-}$ describing
ideal operation deviates from the singlet-state because of the factor $\hat{U%
}_{3}(\theta _{1},\theta _{2})$, which is consistent with Ref. \cite%
{Two-qubit QPT2}. 

To understand these results, we need a decoherence model. Let us first call
our attention to the function of the BS in the polarization space. Let $\hat{%
a}$ and $\hat{b}$ ($\hat{c}$ and $\hat{d}$ ) denote input (output) field
operators of the BS, and the polarization be indicated by subscripts (%
\textit{e.g.}, by $\hat{a}_{H}$ and $\hat{a}_{V}$ for the \textit{H}- and 
\textit{V}-polarized field operators). If we use a right-handed coordinate
system defined by the positive directions of \textit{H}-, \textit{V}-
polarized field ($p$ and $s$ components) and propagation vector (Fig. \ref%
{F1}) to describe the input and output fields, their operator equations are
given by: 
\begin{align}
\hat{c}_{H}& =\sqrt{\mathcal{T}}\hat{a}_{H}+ie^{i(\Gamma +\pi )}\sqrt{%
\mathcal{R}}\hat{b}_{H},  \nonumber \\
\hat{d}_{H}& =ie^{-i(\Gamma +\pi )}\sqrt{\mathcal{R}}\hat{a}_{H}+\sqrt{%
\mathcal{T}}\hat{b}_{H},  \nonumber \\
\hat{c}_{V}& =\sqrt{\mathcal{T}}\hat{a}_{V}+ie^{i\Delta }\sqrt{\mathcal{R}}%
\hat{b}_{V},  \nonumber \\
\hat{d}_{V}& =ie^{-i\Delta }\sqrt{\mathcal{R}}\hat{a}_{V}+\sqrt{\mathcal{T}}%
\hat{b}_{V},  \label{eq7}
\end{align}%
where $\mathcal{T}$ and $\mathcal{R}$ are respectively the transmission and
reflection coefficients of the BS. We have assumed that the $p$ and $s$
components of the field respectively acquire phase shifts $\Gamma +\pi $ and 
$\Delta $ upon reflection. The factor $\pi $ accounts for abrupt change in $p
$-$s$ phase shift due to nonadiabatic changes in the $\mathbf{k}$ vector
upon reflection, which has been frequently noted in the papers concerning the topological
phase effect in optics \cite{Kitano,Berry,Topological phase}. Note that the
helicity of a photon in a state $\left\vert \circlearrowleft \right\rangle $
changes sign upon reflection from the BS but not upon transmission through
the BS if $\Delta -\Gamma =0$. If we consider the output fields that can
contribute to coincidence measurement, we conclude from Eqs. \ref{eq7} that
the BS projects the input state onto $\mathcal{\hat{P}}^{-}$ with $\theta
_{1}=\theta _{2}=\Delta -\Gamma $, which may be finite for a non-polarizing
BS. Note that the projector $\mathcal{P}^{-}$ is the coherent sum of the
identity operator $\hat{U}_{I}$ associated with transmission and the
operator $\hat{U}_{3}(\theta _{1},\theta _{2})\hat{U}_{S}$ associated with
reflection, where $\hat{U}_{3}(\theta _{1},\theta _{2})$ accounts for the
helicity reversal and the extra $p$-$s$ phase shift upon reflection.

Although the above description may suffice when $\tau=0$, the temporal (%
\textit{T}) space of the two-photon state should be taken into account when $%
\tau\neq0$. Concerning the temporal state space, it would be reasonable to
assume that the BS leaves the state untouched on transmission and swaps the
state on reflection. We thus assume that the function of the BS is described
by the following operator entangled with respect to the polarization and
temporal state spaces: 
\begin{equation}
\mathcal{\hat{P}}^{PT}=\mathcal{T}\hat{U}_{I}^{P}\otimes\hat{U}_{I}^{T}%
\mathcal{-R}\hat{U}_{3}^{P}(\theta_{1},\theta_{2})\hat{U}_{S}^{P}\otimes\hat{%
U}_{S}^{T},  \label{eq8}
\end{equation}
where the superscripts indicate the space on which each operator acts. We
further assume that input state is unentangled, \textit{i.e.}, that it is a
product state of polarization state $\hat{\rho}$ and temporal state $\hat{%
\omega}$. Then, by partially tracing over the temporal degree of freedom, we
obtain the following output polarization state $\mathcal{E}(\hat{\rho})$
consistent with Eqs. \ref{eq5} and \ref{eq6}:

\begin{align}
\mathcal{E}(\hat{\rho}) & =Tr_{T}\mathcal{\hat{P}}^{PT}\left( \hat{\rho }%
\otimes\hat{\omega}\right) \left( \mathcal{\hat{P}}^{PT}\right) ^{\dag } 
\nonumber \\
& =(1-p)\mathcal{\hat{P}}^{-}\hat{\rho}\mathcal{\hat{P}}^{-\dag }+p\mathcal{%
\hat{P}}^{+}\hat{\rho}\mathcal{\hat{P}}^{+\dag},  \label{eq9}
\end{align}
where the parameter \textit{p} depends on the overlap of the two-photon
amplitudes contributing to coincidence detection (\textit{r}-\textit{r} or 
\textit{t}-\textit{t}) and thus on the delay $\tau$. Ideally, \textit{p} is
zero if $\tau=0$, increases as $\tau$ increases, and approaches 1/2 when $%
\tau\rightarrow\infty$. Figure \ref{F4} shows that this model largely
explains the experimental results, except that $\theta_{1}$ is not actually
equal to $\theta_{2}$ as in the model. The reason is uncertain, but this
inequality may be due to polarization-dependent $\mathcal{R}/\mathcal{T}$
ignored in the model.

We conclude that the function of the ideal BS is described by $\mathcal{\hat{%
P}}^{-}$, which depends on the $p$-$s$ phase shift as well as the
coefficients $\mathcal{R}$ and $\mathcal{T}$. These parameters obviously
depend on the detailed design of the BS. Properly designed BS would satisfy $%
\mathcal{R}/\mathcal{T}=1$ and $\Delta -\Gamma =0$ simultaneously, at least
for photons that have a particular wavelength and are incident at 45${%
{}^{\circ }}$. In this idealized case, $\mathcal{\hat{P}}^{-}$ agrees with
one of the four Bell states: a triplet state. This shows that the common
belief that an ideal BS behaves as a singlet-state filter is a misconception
due to overlooking the helicity reversal of the photon upon reflection at
the BS.

In summary, we used QPT to investigate the nonideal two-qubit quantum-state
filter and identify its function. We showed that our results are
satisfactorily explained by a simple model with several reasonable
assumptions that a common belief about the function of a BS followed by
post-selection process is mistaken.

We thank S. Ishizaka and A. Tomita for their helpful discussions. This work
was supported by the CREST program of the Japan Science and Technology
Agency.

\end{document}